\pdfoutput=1
\documentclass[10pt, letter, onecolumn]{arxiv}

\usepackage{kantlipsum, lipsum}
\usepackage{dm-colors}
\usepackage{amsmath}
\usepackage{verbatim}
\usepackage{multirow}
\usepackage{scalerel}
\usepackage{booktabs}
\usepackage{enumitem}
\usepackage{xspace}
\usepackage{bm}
\usepackage{bbm}
\usepackage{mathtools}
\usepackage{soul}
\usepackage{epsfig}
\usepackage{graphicx}
\usepackage{amssymb}
\usepackage{colortbl}
\usepackage{csquotes}
\usepackage{setspace}
\usepackage{colortbl}
\usepackage{adjustbox}
\usepackage{tabularx,ragged2e}
\usepackage{placeins}
\usepackage{amsmath}
\usepackage{tcolorbox}
\usepackage{xcolor}
\colorlet{darkgreen}{green!65!black}
\colorlet{darkred}{red!80!black}

\usepackage[symbol]{footmisc}
\usepackage[bibstyle=nature,citestyle=numeric-comp,%
            natbib=true,backend=biber,maxbibnames=99,%
            giveninits=false,sorting=none]{biblatex}
\usepackage{nameref}
\usepackage{varioref}
\usepackage[pagebackref=false,breaklinks=false,%
            colorlinks=true,bookmarks=true,citecolor=ourdarkblue,%
            urlcolor=ourdarkblue,linkcolor=ourdarkblue]{hyperref}
\usepackage[noabbrev,capitalize]{cleveref}
\usepackage{etoc}

\graphicspath{{figures/}}

\title{{\Huge{Towards Accurate Differential Diagnosis with Large Language Models}
}}

\author[$\ast$, $\ddagger$, 1]{Daniel McDuff} 
\author[$\ast$, $\ddagger$, 1]{Mike Schaekermann} 
\author[$\ast$, 1]{Tao Tu} 
\author[$\ast$, 1]{Anil Palepu} 
\author[1]{\\Amy Wang} 
\author[1]{Jake Garrison} 
\author[1]{Karan Singhal} 
\author[1]{Yash Sharma} 
\author[2]{Shekoofeh Azizi} 
\author[1]{\\Kavita Kulkarni} 
\author[1]{Le Hou} 
\author[2]{Yong Cheng} 
\author[1]{Yun Liu} 
\author[1]{\\S Sara Mahdavi} 
\author[1]{Sushant Prakash} 
\author[1]{Anupam Pathak} 
\author[1]{Christopher Semturs} 
\author[1]{\\Shwetak Patel}
\author[1]{Dale R Webster}
\author[1]{Ewa Dominowska} 
\author[1] {\\ Juraj Gottweis}
\author[2]{Joelle Barral}
\author[1]{Katherine Chou}
\author[1]{Greg S Corrado} 
\author[1]{Yossi Matias} 
\author[$\dagger$, $\ddagger$,1]{\\Jake Sunshine} 
\author[$\dagger$, $\ddagger$,1]{Alan Karthikesalingam}
\author[$\dagger$, $\ddagger$,1]{Vivek Natarajan}  

\affil[1]{Google Research, }
\affil[2]{Google DeepMind }

\renewcommand{\correspondingauthor}[1]{$\ast$~Equal contributions. %
                                       $\dagger$~Equal leadership. \\%
                                       $\ddagger$~Corresponding authors: \{dmcduff, mikeshake, jakesunshine, alankarthi, natviv\}@google.com }

\begin{document}
\begin{refsection}

\begin{abstract}
An accurate differential diagnosis (DDx) is a cornerstone of medical care, often reached through an iterative process of interpretation that combines clinical history, physical examination, investigations and procedures. Interactive interfaces powered by Large Language Models (LLMs) present new opportunities to both assist and automate aspects of this process. In this study, we introduce an LLM optimized for diagnostic reasoning, and evaluate its ability to generate a DDx alone or as an aid to clinicians. 20 clinicians evaluated 302 challenging, real-world medical cases sourced from the New England Journal of Medicine (NEJM) case reports. Each case report was read by two clinicians, who were randomized to one of two assistive conditions: either assistance from search engines and standard medical resources, or LLM assistance in addition to these tools. All clinicians provided a baseline, unassisted DDx prior to using the respective assistive tools.  Our LLM for DDx exhibited standalone performance that exceeded that of unassisted clinicians (top-10 accuracy 59.1\% vs 33.6\%, [\textit{p} = 0.04]). 
Comparing the two assisted study arms, the DDx quality score was higher for clinicians assisted by our LLM (top-10 accuracy 51.7\%)
compared to clinicians without its assistance (36.1\%) (McNemar's Test: 45.7, \textit{p} $<$ 0.01) and clinicians with search (44.4\%) (4.75, \textit{p} = 0.03). 
Further, clinicians assisted by our LLM arrived at more comprehensive differential lists than those without its assistance. Our study suggests that our LLM for DDx has potential to improve clinicians' diagnostic reasoning and accuracy in challenging cases, meriting further real-world evaluation for its ability to empower physicians and widen patients' access to specialist-level expertise.
\end{abstract}

\maketitle


\section{Introduction}
\label{sec:introduction}

An accurate diagnosis is a critical component of effective medical care. Building AI systems capable  of performing or assisting clinicians in this important task has been a long-standing grand challenge~\cite{szolovits1978categorical}. While prior focus has been on evaluating a machine's ability to accurately output a diagnosis~\cite{liu2020deep, rauschecker2020artificial, balas2023conversational, kanjee2023accuracy}, real-world clinical practice involves an iterative and interactive process of reasoning about a differential diagnosis (DDx), weighing multiple diagnostic possibilities in the light of increasing amounts of clinical information over time (ranging from clinical history and examination to investigations and procedures). 
Deep learning has been applied to promising effect for generating DDx in a number of specialties including radiology~\cite{rauschecker2020artificial}, ophthalmology~\cite{balas2023conversational} and dermatology~\cite{liu2020deep}, but such systems lack the interactive capabilities to fluently assist a user through communication in natural language.

The emergence of Large Language Models (LLMs) present an opportunity to design novel interactive tools and interfaces to aid in differential diagnosis. Such LLMs trained on vast corpora of text, can recognize, summarize, predict, and generate new text based on knowledge gained during the learning process and task specification via a prompt. These models have demonstrated the ability to perform complex language comprehension and reasoning tasks, generating coherent text and thereby enabling a large variety of real-world applications~\cite{openai2023gpt4,anil2023palm,scao2022bloom,touvron2023llama}.

Both general-purpose LLMs (GPT-4) and medical domain-specialized LLMs (Med-PaLM 2) have demonstrated strong performance in standardized and multiple-choice medical benchmarks~\cite{singhal2023large, nori2023capabilities}. Such evaluations represent a natural starting point for probing the medical knowledge and capabilities but fail to measure utility in real-world scenarios for care delivery, for example in challenging medical cases faced by trained physicians. It is also not obvious how these models might actively assist clinicians in the development of a DDx. Recent work has begun to assess the standalone performance of these models on challenging case reports that involve complex deduction~\cite{kanjee2023accuracy,eriksen2023use,buckley2023accuracy}, but has stopped short of evaluating how they can assist clinicians and augment performance and empower them to provide better care.

In this work, we introduced and investigated the ability of an LLM optimised for clinical diagnostic reasoning, to generate a DDx in challenging, real-world medical cases. Beyond measuring standalone performance like prior work~\cite{kanjee2023accuracy}, we integrated this model into an interactive interface to measure how well our LLM could assist clinicians in developing a DDx. Using a set of challenging real-world cases from the New England Journal of Medicine (NEJM) case reports, we compared clinicians' ability to form a DDx with the assistance of our LLM, versus with access to traditional information retrieval tools (e.g., Internet search and books). The LLM achieved impressive performance in both generating DDx lists that contained the correct diagnosis (i.e., top-10 accuracy) and in identifying the correct final diagnosis as the most likely in the list (i.e., top-1 accuracy). Under automated model based evaluation, the quality and the accuracy of the DDx list produced by our LLM was found to be significantly better than the state-of-the-art GPT-4  model~\cite{kanjee2023accuracy}. 

Perhaps, more importantly, the LLM also improved the diagnostic capability of clinicians as measured by the quality of their DDx lists for the evaluated cases. LLMs optimized for the safety-critical medical domain such as ours present a novel paradigm for assisting clinicians because of the potential for variation in the ways in which a given individual may converse with the system and utilise it in collaborative reasoning. We used semi-structured qualitative interviews to gather information from participating clinicians on their experiences of using the tool, their views of the potential role and risks of LLMs in medical diagnosis and in aiding the differential diagnosis process. These interviews highlighted the potential for LLMs to increase the diversity of DDx lists and speed up the process of arriving at a comprehensive DDx for challenging cases. The clinicians also highlighted that the most appropriate application at the present time would be in learning and education.

Our key contributions can be summarized as:
\begin{itemize}
    \item Introducing an LLM for DDx, a model optimized for differential diagnosis, alongside a user interface allowing clinicians to interact with the model for improving clinical diagnostic reasoning. 
    \item Evaluating the performance of the LLM on challenging diagnostic cases from the NEJM Case Reports.
    \item Showing that the LLM outperforms the prior state of the art, GPT-4, in both top-1 and top-10 accuracy on this benchmark under automated evaluation. 
    \item Evaluating the impact of the LLM as an assistive tool for clinicians in differential diagnosis, with randomized comparison to the usual practice in which clinicians are assisted by Search and their usual clinical resources. 
\end{itemize}

\section{NEJM Clinicopathological Conference Case Reports}

The Case Records of the Massachusetts General Hospital (MGH) are published (lightly edited) transcriptions of the clinicopathological conferences of the MGH (Boston, MA). In the clinicopathological conference, a patient case presentation is described and then an expert physician is asked to provide a DDx and a final diagnosis, along with their diagnostic reasoning, based only on the patient's provided medical history and preliminary test results. The published cases, organized generally as diagnostic puzzles culminating in a definitive, pathology-confirmed diagnosis, are published regularly in the NEJM.  We leverage these case reports, licensed from the NEJM, to evaluate the LLM’s capability to generate a DDx alone and, separately, to aid clinicians in generation of their own differential. For this latter task, we developed a user interface for clinicians to interact with the LLM.

A set of 326 case texts from the NEJM Clinicopathological Conference (CPC) series were considered. These case reports were published over a 10 year period between June 13\textsuperscript{th} 2013 and August 10\textsuperscript{th} 2023. Of these, 23 (7\%) were excluded on the grounds that they discussed case management and were not primarily focused on diagnosis. The articles were distributed over the years between 2013-2023 as follows: 2013 N=22, 2014 N=34, 2015 N=36, 2016 N=35, 2017 N=36, 2018 N=16, 2020 N=23, 2021 N=36, 2022 N=39, 2023 N=26. The supplementary material includes the full set of case numbers. The 302 cases include the 70 cases used by Kanjee et al.~\cite{kanjee2023accuracy}. 

These case reports cover a range of medical specialties. The largest proportion are from internal medicine (N=159), followed by neurology (N=42), pediatrics (N=33) and psychiatry (N=10). The text corresponding to the history of the present illness (HPI) was manually extracted from each article as input to the LLM. The average (median) word count of these sections of the case reports is 1,031 words (mean: 1,044, SD: 296, range: 378-2,428). The average (median) character count is 6,619 characters (mean: 6,760, SD: 1,983, range: 2,426-15,196). 

\begin{figure*}[t]
    \begin{center}
    \includegraphics[width=1\textwidth]{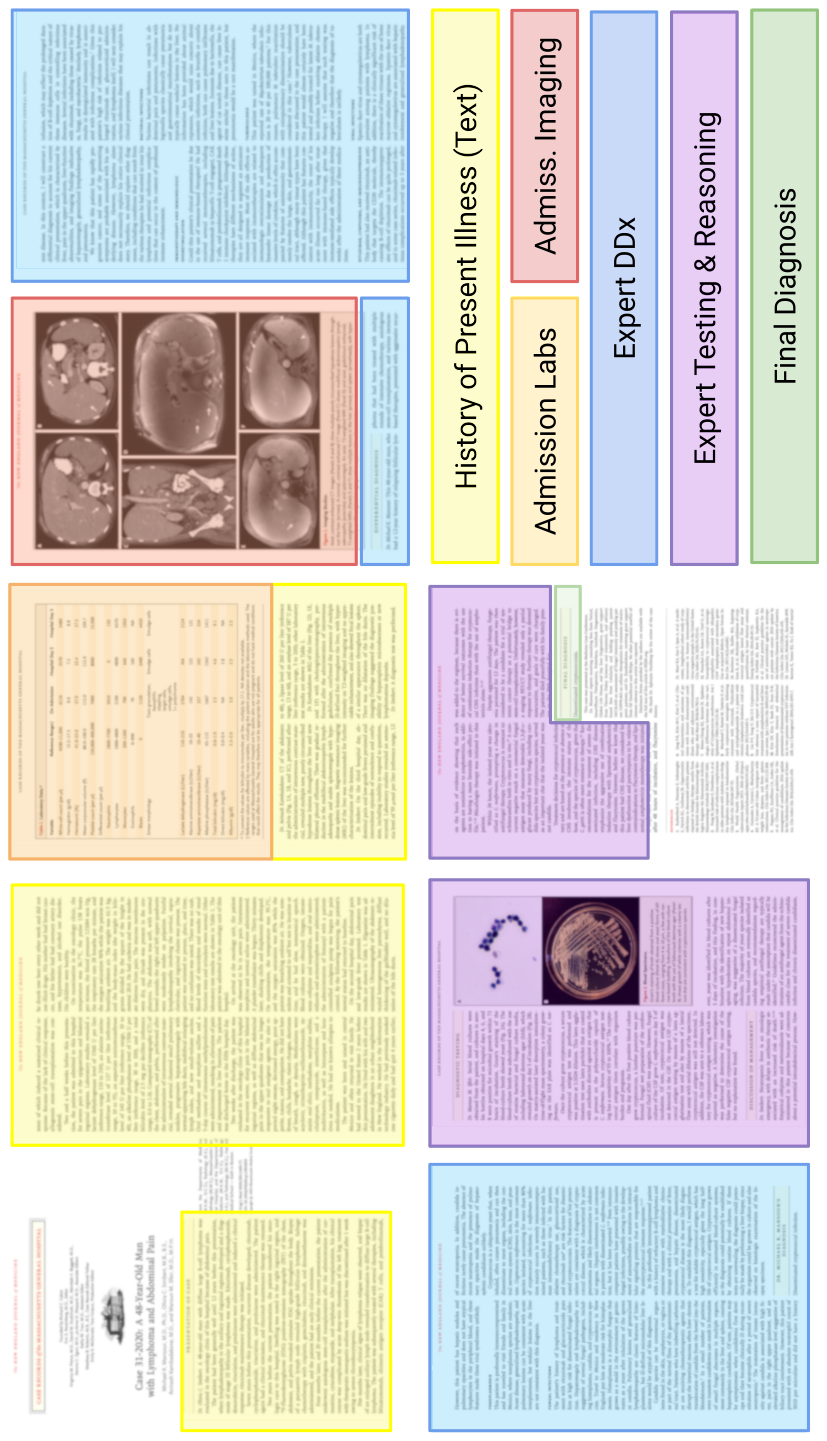}
    \end{center}
    \caption{\textbf{NEJM Clinicopathological Conference Case Reports.} History of Present Illness, Admission Labs and Admission Imaging sections were included in the redacted version presented to generalist clinicians for producing a DDx. The LLM had access to only the History of Present Illness. Specialist clinicians evaluating the quality of the DDx had access to the full (unredacted) case report including the expert differential discussion.}
    \label{fig:case_explanation}
\end{figure*}

A modified version of the article, inclusive of the provided HPI, admission imaging and admission labs (if available in the case) was created for the human clinicians (see Fig.~\ref{fig:case_explanation}). This version had redacted the final diagnosis, expert discussion of the DDx and any subsequent imaging or biopsy results (which are typical elements of the conclusion of the case challenges). Given the LLM is a text-only AI model, the admission images and lab tables were not fed into the model. However, text-based descriptions of specific lab values or imaging findings were sometimes included in the case description.

\section{Training a Large Language Model for DDx}
Our study introduces an LLM for DDx, a model which uses a transformer architecture (PaLM 2~\cite{anil2023palm}), fine-tuned on medical domain data; alongside an interface for enabling its use as an interactive assistant for clinicians.

As with Med-PaLM 2~\cite{singhal2023large}, our LLM builds upon PaLM 2, an iteration of Google’s LLM with substantial performance improvements on multiple LLM benchmark tasks. For the purposes of this analysis the large (L) PaLM 2 model was used.

The LLM was fine-tuned with long context length on a task mixture consisting of medical question answering (multiple-choice and long-form questions), medical dialogue generation and electronic health record (EHR) note summarization. The datasets used included the training splits of MultiMedQA (MedQA, MedMCQA, HealthSearchQA, LiveQA and MedicationQA)~\cite{singhal2023large}, a proprietary dataset of medical conversations, and expert handcrafted EHR note summaries from MIMIC-III~\cite{johnson2016mimic}. The capability to process long context input enables the LLM to handle tasks that require long-range reasoning and comprehension.

\textbf{Zero-Shot Prompting.} We evaluated the LLM on each of the NEJM case studies with the following prompt: ``You are a helpful medical assistant. You will be provided and asked about a complicated clinical case; read it carefully and then provide a diverse and thorough DDx". 

\textit{Example Input:}
\begin{tcolorbox}

\textbf{Case:}  \color{red} [Case Text]\color{black} 
\\
What are the top 10 most likely diagnoses? Be precise, listing one diagnosis per line, and try to cover many unique possibilities (at least 10). The top 10 diagnoses are:

\end{tcolorbox}

\textit{Example Output:}
\begin{tcolorbox}

\begin{enumerate}
    \item Sézary syndrome
    \item Adult T-cell leukemia/lymphoma
    \item Mycosis fungoides
    \item Atopic dermatitis
    \item Psoriasis
    \item Drug reaction with eosinophilia and systemic symptoms (DRESS)
    \item Graft-versus-host disease (GVHD)
    \item Cutaneous T-cell lymphoma, not otherwise specified
    \item Hypereosinophilic syndrome
    \item Systemic lupus erythematosus (SLE)
\end{enumerate}

\end{tcolorbox}

\section{The LLM for DDx User Interface} 
\begin{figure*}[t]
    \begin{center}
    \includegraphics[width=1.0\textwidth]{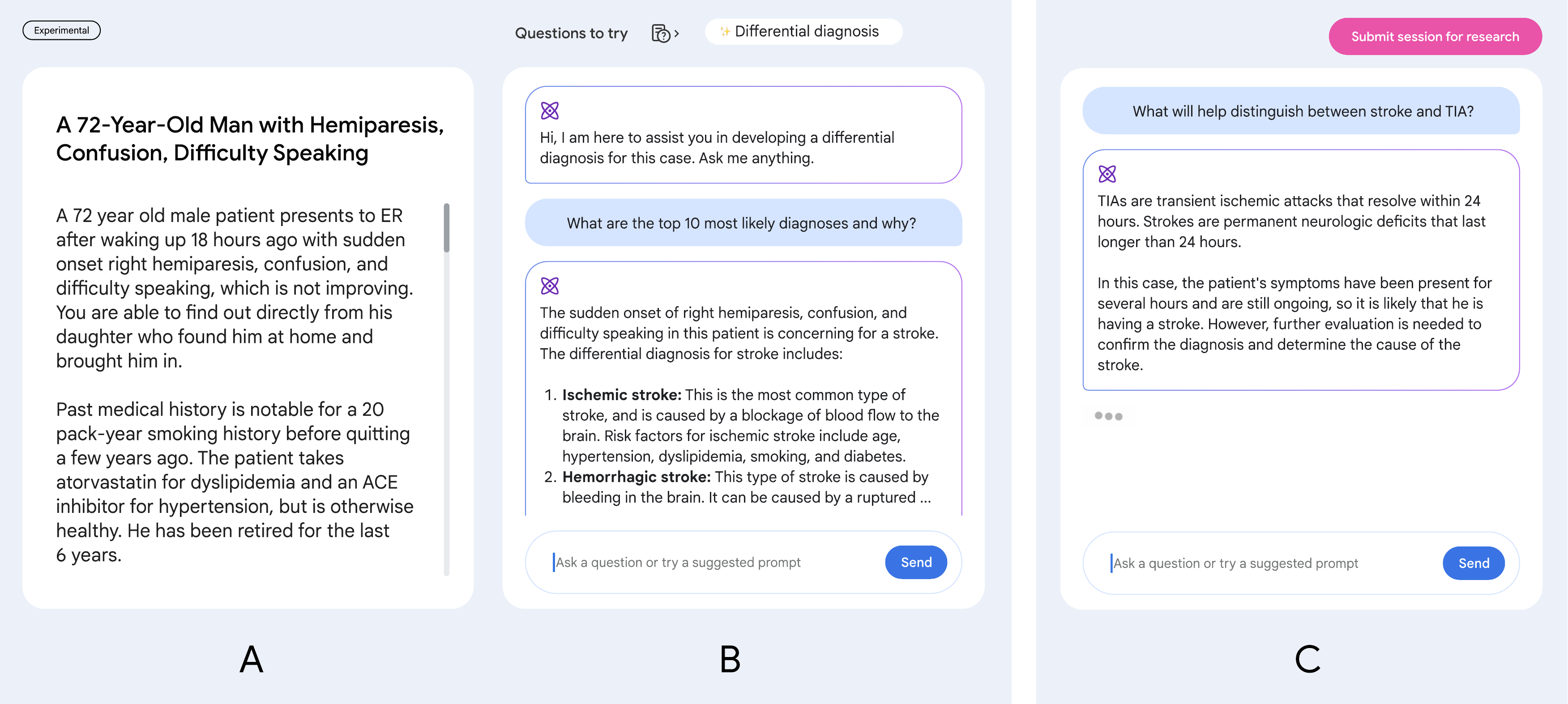}
    \end{center}
    \caption{\textbf{The LLM for DDx User Interface.} The history of the present illness (text only) was pre-populated in the user interface (A) with an initial suggested prompt to query the LLM (B). Following this prompt and response, the user was free to enter any additional follow-up questions (C). The case shown in this figure is a mock case selected for illustrative purposes only.}
    \label{fig:user_interface}
\end{figure*}

The interface associated with our LLM, depicted in Fig.~\ref{fig:user_interface}, enables users to interact with the underlying model via text-based chat in the context of a given case description.
In our study, the interface was pre-populated with a text-only representation of the history of present illness (HPI) for a given case.
Clinicians were asked to initiate the interaction by querying the LLM using a suggested prompt. Following this initial prompt and the LLM's response, clinicians were free to query the model using any additional follow-up questions, though clinicians were cautioned to avoid asking questions about information that had not already been presented in the case. A pilot study indicated that without such a warning, clinicians may ask questions about specific lab values or imaging leading to confabulations.

For a given question, the interface generated the response by querying the LLM using prompt template:

\begin{tcolorbox}

Read the case below and answer the question provided after the case. \\
Format your response in markdown syntax to create paragraphs and bullet points. Use `<br><br>' to start a new paragraph. Each paragraph should be 100 words or less. Use bullet points to list multiple options. Use `<br>* ' to start a new bullet point. Emphasize important phrases like headlines. Use `**' right before and right after a phrase to emphasize it. There must be NO space in between `**' and the phrase you try to emphasize. \\
\textbf{Case:} \color{red}[Case Text]\color{black}  \\
\textbf{Question (suggested initial question is ``What are the top 10 most likely diagnoses and why (be precise)?''):} \color{blue} [Question]\color{black}  \\
\textbf{Answer:}
\end{tcolorbox}

\section{Methods}

\subsection{Experimental Design}

\begin{figure*}[t]
    \begin{center}
    \includegraphics[width=1\textwidth]{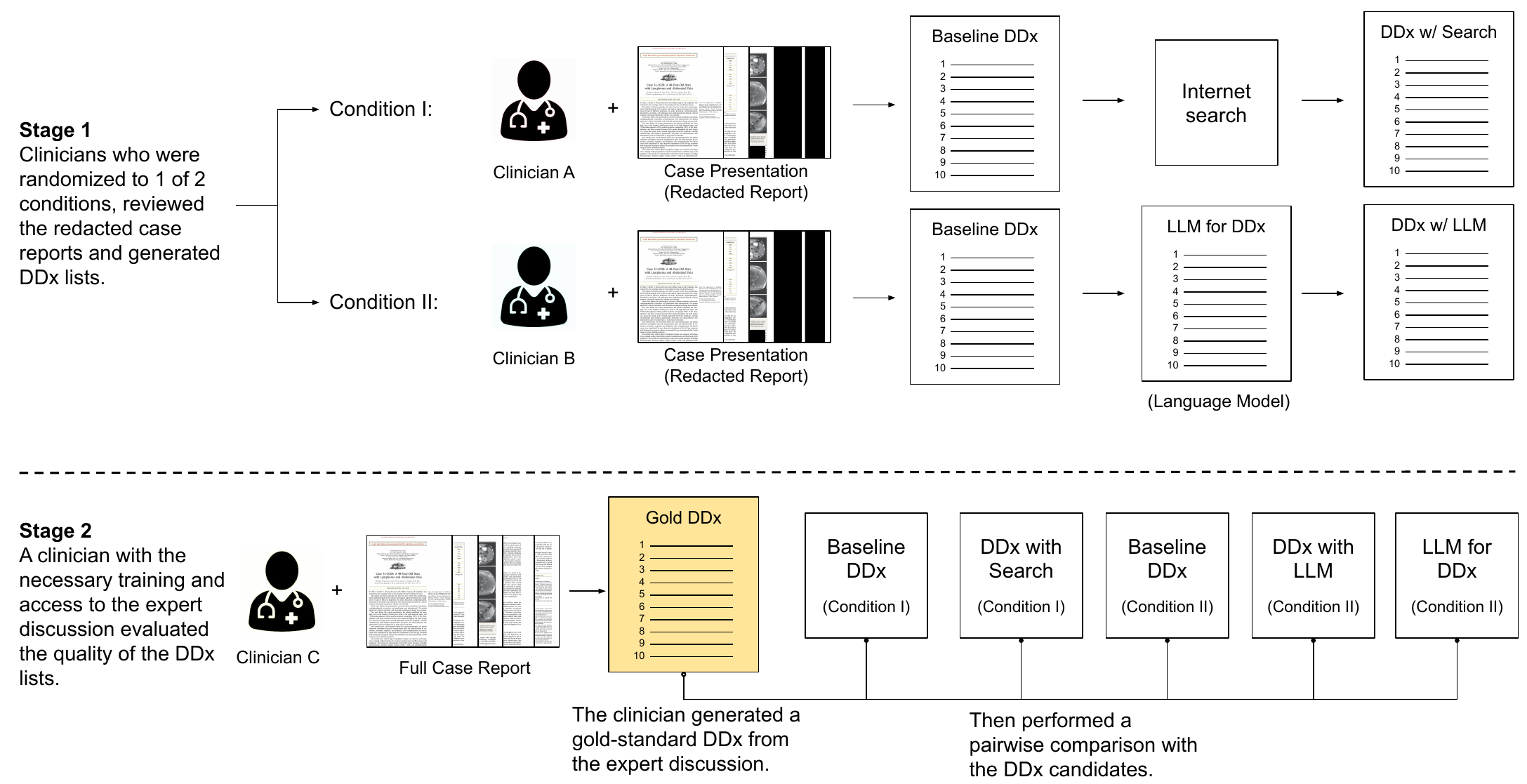}
    \end{center}
    \caption{\textbf{Experimental Design.} To evaluate the LLM's ability to generate DDx lists and aid clinicians with their DDx generation, we designed a two-stage reader study. First, clinicians with access only to the case presentation completed DDx lists without using any assistive tools. Second, the clinicians completed DDx lists with access either to Search engines and other resources, or to LLM in addition to these tools. Randomization was employed such that every case was reviewed by two clinicians, one with LLM assistance and one without. These DDx lists were then evaluated by a specialist who had access to the full case and expert commentary on the differential diagnosis, but who was blinded to whether and what assistive tool was used.}
    \label{fig:experimentaldesign}
\end{figure*}

In order to comparatively evaluate the LLM's ability to generate a DDx alone and aid clinicians with their DDx generation we designed a two-stage reader study illustrated in Fig.~\ref{fig:experimentaldesign}). Our study was designed to evaluate the assistive effect of the LLM for generalist clinicians (not specialists) who only have access to the case presentation and not the full case information (which would include the expert commentary on the DDx).
The first stage of the study had a counterbalanced design with two conditions. Clinicians generated DDx lists first without assistance and then a second time with assistance, where the type of assistance varied by condition.

\textbf{Stage 1. Clinicians generate DDx with and without assistance}

Twenty U.S. board-certified internal medicine physicians (median years of experience: 9, mean: 11.5, SD: 7.24, range: 3-32) viewed the redacted case report, with access to the case presentation and associated figures and tables.  They did this task in one of two conditions, based on random assignment. 

\textit{Condition I - Search.} The clinicians were first instructed to provide a list of up to ten diagnoses, with a minimum of three, based solely on review of the case presentation without using any reference materials (e.g., books) or tools (e.g., Internet search). Following this, the clinicians were instructed to use Internet Search or other resources as desired (but not given access to the LLM) and asked to re-perform their DDx.

\textit{Condition II - LLM for DDx.} As with condition I, the clinicians were first instructed to provide a list of up to 10 diagnoses, with a minimum of three, based solely on review of the case presentation without using any reference materials (e.g., books) or tools (e.g., Internet search). Following this the clinicians were given access to the LLM and asked to re-perform their DDx. In addition to the LLM, clinicians could choose to use Internet search or other resources if they wished.

\textbf{Stage 2. Specialists with full case information extract gold DDx and evaluate Stage 1 DDx}

Nineteen U.S. board-certified specialist clinicians (median years of experience: 14, mean: 13.7, SD: 7.82, range: 4-38) were recruited from internal medicine (N=10), neurology (N=3), pediatrics (N=2), psychiatry (N=1), dermatology (N=1), obstetrics (N=1), and emergency medicine (N=1). Their mean years of experience was 13.7 (SD: 7.82, range: 4-38). These specialists were aligned with the specialty of the respective CPC case, viewed the full case report and were asked to list at least 5 and up to 10 differential diagnoses. 
Following this, they were asked to evaluate the five DDx lists generated in Stage 1, including two DDx lists from condition 1 (DDx without Assistance and DDx with Search Assistance), two DDx lists from condition 2 (DDx without Assistance and DDx with LLM Assistance) and the standalone LLM DDx list.

The specialists answered the following questions to evaluate the DDx lists: 

The \textbf{quality score} developed by Bond et al.~\cite{bond2012differential} and used by Kanjee et al.~\cite{kanjee2023accuracy} is a differential score based on an ordinal 5-point scale: 
``How close did the differential diagnoses (DDx) come to including the final diagnosis?''. The options were: 5. DDx includes the correct diagnosis, 4. DDx contains something that is very close, but not an exact match to the correct diagnosis, 3. DDx contains something that is closely related and might have been helpful in determining the correct diagnosis, 2. DDx contains something that is related, but unlikely to be helpful in determining the correct diagnosis, 1. Nothing in the DDx is related to the correct diagnosis.

An \textbf{appropriateness score}:
``How appropriate was each of the differential diagnosis lists from the different medical experts compared the differential list that you just produced?''. The options to respond were on a Likert scale of 5 (very appropriate) to 1 (very inappropriate).

A \textbf{comprehensiveness score}:
``Using your differential diagnosis list as a benchmark/gold standard, how comprehensive are the differential lists from each of the experts?''. The options to respond were: 4. The DDx contains all candidates that are reasonable, 3. The DDx contains most of the candidates but some are missing, 2. The DDx contains some of the candidates but a number are missing, 1. The DDx has major candidates missing.

Finally, specialists were asked to specify in which position of the DDx list the correct diagnosis was matched, in case it was included in the DDx at all.

\textbf{Automated Evaluation.}
In addition to comparing against ground-truth diagnosis and expert evaluation from clinicians, we also created an automated evaluation of the performance of the five DDxs using a language-model based metric. Such automated metrics are useful as human evaluation is time and cost-prohibitive for many experiments. We first extracted the (up to ten) individual diagnoses listed in each DDx. We leveraged minor text-processing steps via regular expressions to separate the outputs by newlines and strip any numbering before the diagnoses. Then we asked a medically fine-tuned language model, Med-PaLM 2~\cite{singhal2023large}, whether or not each of these diagnoses was the same as the ground-truth diagnosis using the following prompt:

\begin{tcolorbox}
Is our predicted diagnosis correct (y/n)? Predicted diagnosis: \color{red}[diagnosis]\color{black}, True diagnosis: \color{red}[label]\color{black}

Answer [y/n].
\end{tcolorbox}

 A diagnosis was marked as correct if the language model output 'y'.

\subsection{Qualitative Interviews}

Following the study we performed a semi-structured 30-minute interviews with five of the generalist clinicians who participated in Stage 1. Semi-structured interviews explored the following questions:  

\begin{enumerate}
    \item How did you find the task of generating a DDx from the case report text?
    \item Think about how you used Internet search or other resources. How were these tools helpful or unhelpful?
    \item Think about how you used the LLM for DDx. How was it helpful or unhelpful?
    \item Were there cases where you trusted the output of the search queries? Tell us more about the experience if so, such as types of cases, types of search results.
    \item Were there cases where you trusted the output of the LLM queries? Tell us more about the experience if so, such as types of cases, types of search results.
    \item Think about the reasoning provided by the LLM's interface? Where were they helpful?  Where were they unhelpful?
    \item What follow-up questions did you find most helpful to ask the LLM?
    \item How much time does it take to get used to the LLM? How was it intuitive? How was it unintuitive?
\end{enumerate}

\section{Results}

In evaluating the quality of the DDx lists we used several criteria, inspired by the approach taken in~\cite{kanjee2023accuracy} and extended to draw additional insight from the clinicians. First, we measured whether the final diagnosis matched an entry in the DDx list and in which position (top-N accuracy). Second, we used Bond et al.'s~\cite{bond2012differential} quality score and the appropriateness and comprehensiveness scales that we created. Combined these measures assess overall DDx quality, appropriateness and comprehensiveness.

When using the LLM for assistance, clinicians asked, on average (mean), 2.92 questions in the interface (median, 2 and IQR, 1-4).
On average (mean), clinician questions consisted of 9.39 words (median, 10 and IQR, 6-12) and 54.31 characters (median, 61 and IQR, 39-63).
The LLM responses, on average (mean), consisted of 237.60 words (median, 198 and IQR, 127-332) and 1540.81 characters (median, 1276 and IQR, 815-2210).

In the Search condition the most popular tools were UpToDate (used in 34\% of tasks), Google Search (30\%) and PubMed (22\%).  While clinicians were allowed to use additional tools in the LLM condition, this was far less frequent ($<$5\% of tasks).

\subsection{Performance of the Language Model on Differential Diagnosis}

\begin{figure*}[t]
    \begin{center}
    \includegraphics[width=0.99\textwidth]{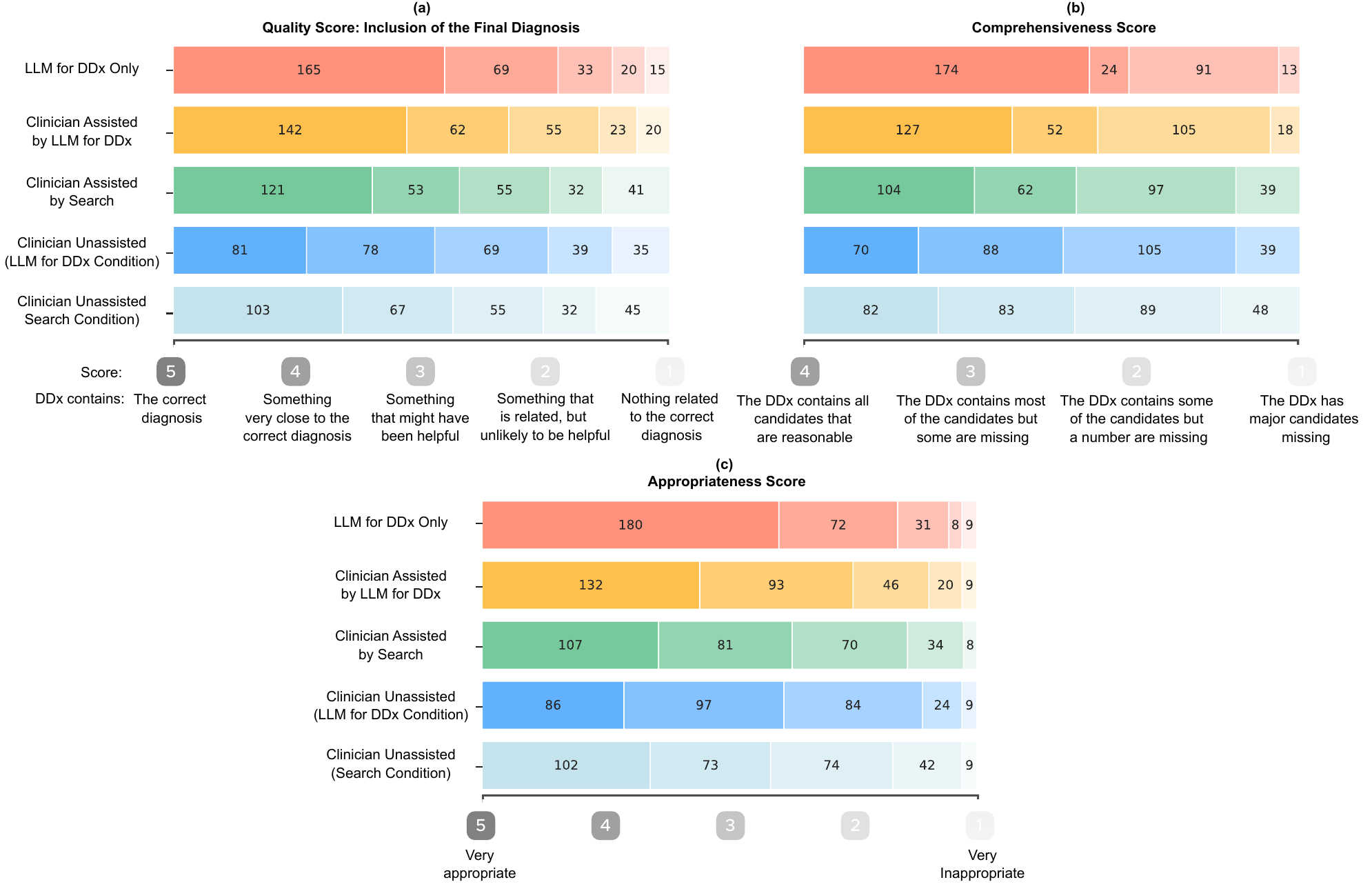}
    \end{center}
    \caption{\textbf{Evaluation of the quality of generalists' DDx lists.} (a) DDx Quality Score based on the question: ``How close did the differential diagnoses (DDx) come to including the final diagnosis?'' (b) DDx Comprehensiveness Score based on the question: ``Using your differential diagnosis list as a bench mark/gold standard, how comprehensive are the differential lists from  each of the experts?'' (c) DDx Appropriateness Score based on the question: ``How appropriate was each of the differential diagnosis lists from the different medical experts compared the differential list that you just produced?''
    In all cases, the LLM and clinicians assisted by the LLM scored the highest overall. Numbers reflect the number of cases (out of 302). Note: The clinicians could answer ``I am not sure'' in response to these questions, in a very small number ($<$ 1\%) of cases they used that option. }
    \label{fig:stacked-bar}
\end{figure*}

\begin{figure*}[t]
    \begin{center}
    \includegraphics[width=0.9\textwidth]{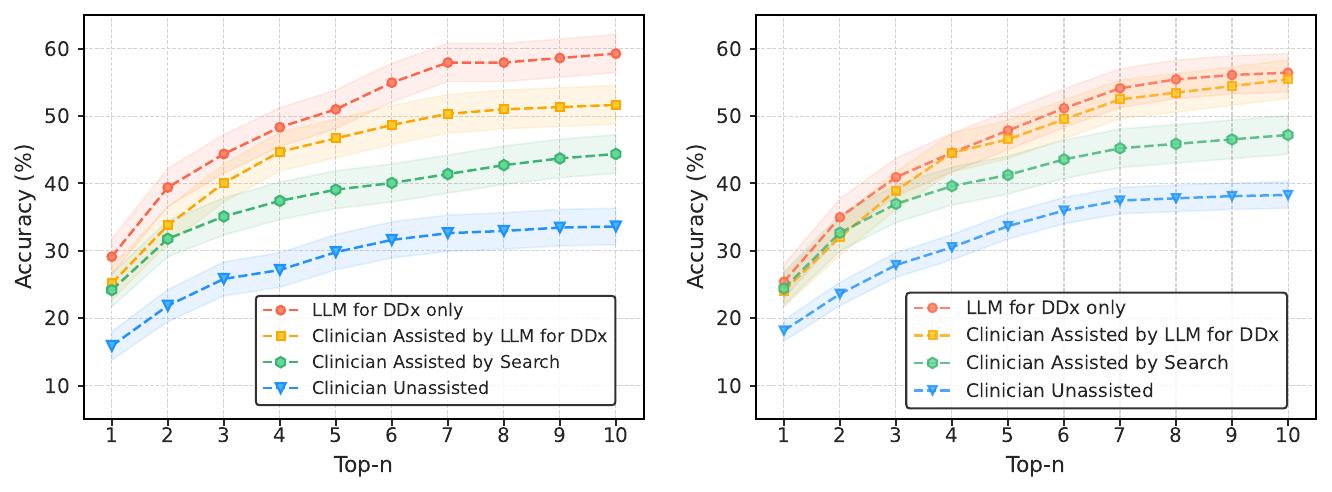}
    \end{center}
    \caption{\textbf{Top-n Accuracy.} (left) The percentage of DDx lists with the final diagnosis through human evaluation. (right) The percentage of DDx lists with the final diagnosis through automated evaluation.}
    \label{fig:top_n_accuracy}
\end{figure*}

\textbf{Quality, Appropriateness and Comprehensiveness.}

Our language model's DDx lists achieved strong quality, appropriateness and comprehensiveness scores (see Fig.~\ref{fig:stacked-bar}). 
The median quality score was 5 (``DDx includes the correct diagnosis'') with 54\% of DDx lists achieving that score.  
The number of cases that scored 5 (i.e., the DDx included the top diagnosis) was statistically significantly higher for the LLM compared to clinicians without assistance (McNemar’s Test: 64.4, \textit{p} $<$ 0.01).  The mean appropriateness score of 4.43 out of five (SD: 0.92). The median comprehensiveness score was 4 (= ``The DDx contains all candidates that are reasonable'') with 55\% of the DDx lists achieving that score.

The mean appropriateness score of the LLM (4.34) was significantly higher than that for unassisted clinicians (3.74) (paired t-test 8.52,\textit{p} $<$ 0.001) 
and assisted clinicians in either the Search (3.80) (paired t-test 7.23, \textit{p} $<$ 0.001) or LLM (4.06) (paired t-test 4.98, \textit{p} $<$ 0.001) conditions.

\textbf{Top-N Accuracy.}
For computing top-N accuracy, if any of the first N diagnoses in an individual DDx were marked correct by the language model, the differential was considered to be correct. We computed the proportion of correct DDx lists across all cases to compute the top-N accuracy (for N from 1 to 10) for each DDx. The LLM reliably generated DDx lists that perform well against the ground-truth diagnosis. Fig.~\ref{fig:top_n_accuracy} shows the top-N accuracy for the LLM. The LLM provided the correct diagnosis in 177 (59\%) of the DDx lists and in 89 (29\%) of the lists it was at the top of the list.  These scores are above the scores the clinicians achieved in any of the conditions. The top-10 accuracy of the LLM (59.1\%) was significantly higher than the top-10 accuracy for the unassisted clinicians (33.6\%) (\textit{p} = 0.04).

Fig.~\ref{fig:top_n_accuracy} shows the top-N accuracy based on human and the automated metric. The results are broadly comparable, illustrating that despite the final diagnoses often being complex and nuanced, the automated metric faithfully captures the distinction between a DDx list that includes the correct diagnosis and one that does not.

\subsection{LLM for DDx as an Assistant for Differential Diagnosis}

\textbf{Quality, Appropriateness and Comprehensiveness.}
Of the DDx lists created before assistance 37\% (Search condition) and 29\% (LLM for DDx condition) achieved a quality score of 5 (Fig.~\ref{fig:stacked-bar}). 
In comparison 49\% of those created with assistance from the LLM scored 5.

The number of cases that scored 5 (i.e., the DDx included the top diagnosis) was statistically higher for clinicians assisted by the LLM compared to clinicians without assistance (McNemar’s Test: 48.3, \textit{p} $<$ 0.01) and clinicians with Search assistance (5.45, \textit{p} = 0.02). 

For comprehensiveness, the number of cases that scored 4 (i.e., The DDx contains all candidates that are reasonable) was statistically higher for clinicians assisted by the LLM compared to clinicians without assistance (McNemar’s Test: 185.8, \textit{p} $<$ 0.01) and clinicians with Search assistance (185.8, \textit{p} $<$ 0.01).

The mean appropriateness score after assistance with the LLM (4.06) was significantly higher than after assistance with Search (3.80) (paired t-test 3.32, \textit{p} = 0.001) and the baseline (3.74) (paired t-test 4.79, \textit{p} $<$ 0.001).

To summarize, with the support of the LLM, the quality, appropriateness and comprehensiveness scores for the DDx lists were greater than for the lists prior to assistance (see Fig.~\ref{fig:stacked-bar}).

\begin{figure*}[t]
    \begin{center}
    \includegraphics[width=1\textwidth]{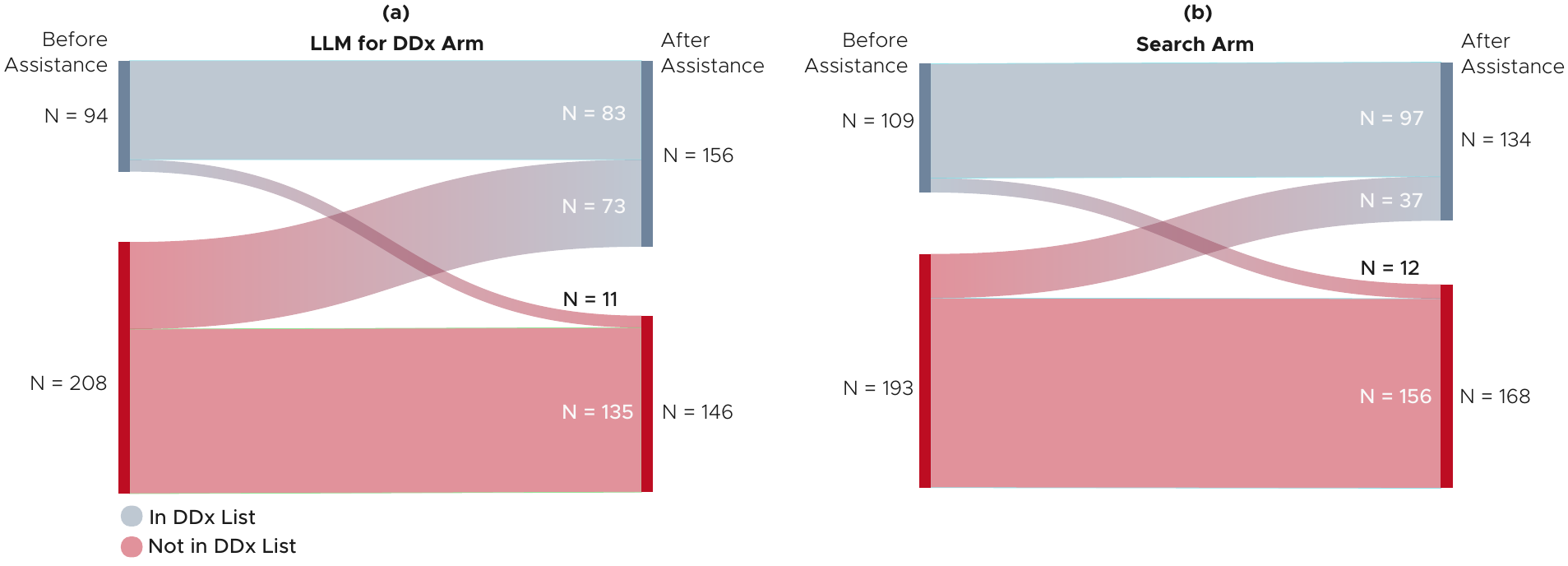}
    \end{center}
    \caption{\textbf{Sankey diagram showing effect of assistance.} (a) in the LLM arm, 73 cases had the final diagnosis in the DDx list \emph{after} assistance that did not contain it \emph{before}, (b) in the Search arm this was 37 cases. In both cases, a small minority (LLM for DDx arm = 11, Search arm = 12) a DDx list with the final diagnosis before assistance did not contain it afterwards.}
    \label{fig:sankey_search}
\end{figure*}

\textbf{Top-N Accuracy.}
The top-N accuracy of the clinicians increased with assistance from the LLM compared to without (see Fig.~\ref{fig:top_n_accuracy}). A Sankey diagram illustrates the impact of the two forms of assistance (Search and LLM for DDx) on top-10 accuracy (Fig.~\ref{fig:sankey_search}). In the LLM condition 73 cases that did not feature the final diagnosis prior to using the tool included it after assistance from the LLM.  This result is in contrast to only 37 cases in the Search condition.
Comparing the two assisted study arms, the DDx quality score was higher for clinicians assisted by our LLM (top-10 accuracy 51.7\%)
compared to clinicians without its assistance (36.1\%) (McNemar's Test: 45.7, \textit{p} $<$ 0.01) and clinicians with search (44.4\%) (4.75, \textit{p} = 0.03).

\subsection{Duration of DDx Tasks with the LLM for DDx and Search}
The time taken to generate updated DDx lists in the Search conditions vs the LLM condition were similar (Search: [7.19 minutes, SD = 5.33], LLM for DDx: [7.29 minutes, SD=6.41]). These were not significantly different (paired t-test \textit{p} = 0.807), which is surprising as the clinicians all had experience using Internet search and other information retrieval tools, yet they were using the LLM interface for the first time.  We hypothesized that they would take longer using the LLM due to the initial learning curve.

\subsection{Length of DDx Lists Using the LLM for DDx and Search}
When unassisted, the median length of the DDx lists was 6 (IQR, 5-9); the mean was 6.41 (SD, 2.39).  With search the median DDx list length was 7 (IQR, 5-10); the mean was 6.92 (SD, 2.52). With the LLM the median DDx list length was 8 (IQR, 6-10); the mean was 7.58 (SD, 2.33).
With assistance from the LLM, the length of the DDx lists was longer than without assistance (paired t-test: 7.13, \textit{p} $<$ 0.001) and longer than the DDx lists with assistance from search (paired t-test: 3.15, \textit{p} = 0.002).

\subsection{Contamination Analysis}

We trained the LLM by building on an model pretrained on large-scale data and fine-tuning on medical data. While we did not include NEJM case reports in the fine-tuning data for the model, it is possible that pretraining data for the model contained partial mentions or full NEJM case reports, whether from the original source (NEJM) or reproduced by other websites. To better understand the possibility of overlap between the training corpus and the test data, we performed a contamination analysis using the pretraining corpora. We looked for overlap between character sequences in the test articles and training corpora using a sliding window, searching for all instances of 512-character overlap. A case report is considered to have overlap if at at least one document from the pretraining corpora has an overlap. We identified that there was no overlap for case reports beginning in 2022 and 2023.  
Some overlap existed for case reports published prior to 2022. We calculated the top-N accuracy for the LLM on both of these sets of case reports, prior to 2022 (N = 238) and 2022 to date (N = 65), and did not observe a substantial difference in results. Across all years, 16.9\% (51 out of 302) of case reports displayed at least one instance of overlap.

\begin{figure*}[t]
    \begin{center}
    \includegraphics[width=0.45\textwidth]{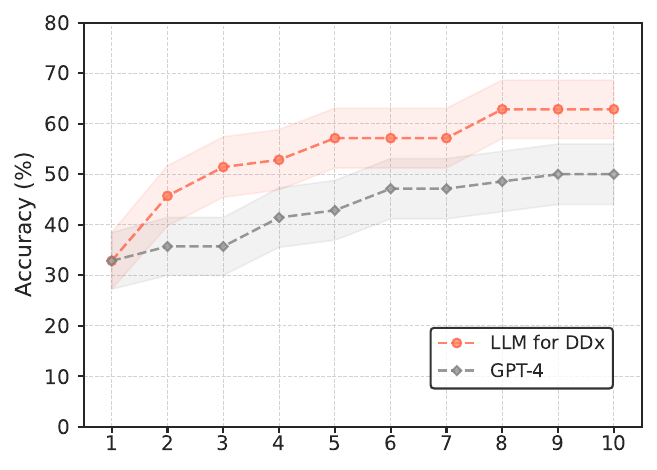}
    \end{center}
    \caption{\textbf{Top-n Accuracy.} Comparison of the percentage of DDx lists with the final diagnosis for our LLM for DDx vs GPT-4 for 70 cases.}
    \label{fig:top_n_accuracy_gpt}
\end{figure*}

\subsection{LLM for DDx Comparison with GPT-4}

As we did not have the same set of human raters who evaluated the differentials produced by GPT-4~\cite{kanjee2023accuracy} and our LLM, we can not compare top-10 accuracy numbers directly. Therefore, in our study design, we evaluate performance on that 70 cases subset (reported in~\cite{kanjee2023accuracy}) using the automated metric (which is shown above to be relatively consistent with human evaluation). Our LLM for DDx performs better with regard to top-N accuracy for N>1, with the gap being most prominent N>2 (Fig.~\ref{fig:top_n_accuracy_gpt}). This suggests potentially significant improvements in quality and comprehensiveness of the differentials produced by our LLM.

\subsection{Qualitative Analysis}

We describe our qualitative results that provide insight into how the clinicians viewed the LLM. We identified several key themes in the clinicians' responses and present several illustrative quotes below. 

\textbf{Comparing Search and the LLM for DDx.}
One clinician contrasted the use of Search to the LLM in this way: \textit{``Search was adequate when I had a good idea of what the differential ought to be to start with, but there were some cases where I was only able to generate 3 or 4 because I really wasn't sure. If I put in `infectious causes of headache' [to the search engine] those were not very helpful.''}, whereas \textit{``[the LLM] was required to pull some additional diagnoses that may not have been the final diagnosis but would be important to think about.''}. 

\textbf{Use Cases.}

C3: \textit{``I think if [the LLM] is intended for medical students or for medical education then it could be very very powerful tool.''}.

\textbf{Confabulations.}

C1: \textit{``I walked into it thinking I could ask what ever I want, but if it was something that could not be inferred from the case I might get a response that isn't real or appropriate.}
C3: \textit{``The biggest thing I think that I had a concern about was inaccuracy for someone who does not have a clinical background''}. But the clinicians found ways to work around these limitations by leveraging their expertise or other resources to validate responses.

\textbf{Ease of Use.}

C2: \textit{``For me it was very intuitive, it was a very easy tool to use, I wish I had it every single day.''}). These comments highlight that natural language interfaces have a very low barrier to use.

\begin{table}[!t]
\setlength{\tabcolsep}{3pt}
\caption{\textbf{Top-1 and Top-10 Accuracy.} The percentage of DDx lists with the final diagnosis.
}
\vspace{-8pt}
\label{exp:table:finetune}
\small
\begin{center}
\adjustbox{max width=\textwidth}{
\begin{tabular}{lcccccccc}
\toprule[1.5pt]
& \multicolumn{2}{c}{Model-Only} & \multicolumn{6}{c}{Human} \\
& \multicolumn{2}{c}{LLM for DDx} & \multicolumn{2}{c}{Before Assistance} & \multicolumn{2}{c}{After Search Assistance} & \multicolumn{2}{c}{After LLM for DDx Assistance} \\
\cmidrule(lr){2-3} \cmidrule(lr){4-5} \cmidrule(lr){6-7} \cmidrule(lr){8-9} 
Metrics & Top-1$^\uparrow$ & Top-10$^\uparrow$ & Top-1$^\uparrow$ & Top-10$^\uparrow$ & Top-1$^\uparrow$ & Top-10$^\uparrow$ & Top-1$^\uparrow$ & Top-10$^\uparrow$  \\ \midrule
Full Set (302 Cases) &  29.2\% & 59.1\% & 15.9\% & 33.6\% & 24.3\% & 44.5\% & 25.2\% & 51.8\% \\[1.2pt]
Set without Overlap (56 Cases)  & 35.4\% & 55.4\% & 13.8\% & 34.6\% &  29.2\% & 46.2\% & 24.6\% & 52.3\% \\[1.2pt]
\midrule
Difference & \textcolor{darkgreen}{\texttt{+}\textbf{6.2\%}} & \textcolor{darkred}{\texttt{-}\textbf{3.7\%}} & \textcolor{darkred}{\texttt{-}\textbf{2.1\%}} & 
\textcolor{darkgreen}{\texttt{+}\textbf{1.0\%}} & 
\textcolor{darkgreen}{\texttt{+}\textbf{4.9\%}} &
\textcolor{darkgreen}{\texttt{+}\textbf{1.7\%}} &
\textcolor{darkred}{\texttt{-}\textbf{0.6\%}} &
\textcolor{darkgreen}{\texttt{+}\textbf{0.5\%}} \\
\bottomrule[1.5pt]
\end{tabular}
}
\end{center}
\vspace{-0.2cm}
\end{table}

\section{Discussion}
We used a popular series of complex diagnostic challenges to evaluate an LLM optimized for clinical reasoning and diagnosis (LLM for DDx); both in a standalone capacity and under randomized comparisons as an assistive tool for physicians. In standalone performance, the LLM generated more appropriate and comprehensive DDx lists than physicians when they were unassisted, with its DDx lists more likely to include the final diagnosis than DDx lists from a board-certified internal medicine physician, no matter which position in the DDx list was considered (i.e., top-N accuracy with N ranging from 1 to 10). Clinicians using the LLM as an assistant produced a DDx with higher top-N accuracy, and DDx with greater quality, appropriateness and comprehensiveness; compared to the status quo for clinical practice (use of Internet search and other resources).

The NEJM clinicopathological conferences (CPCs) examined here are well-known for being unique and challenging clinical conundrums. Within this distinctive setting, the proposed LLM outperformed an unassisted board-certified physician in both top-1 and top-n performance. While the CPCs have long been used as benchmarks for difficult diagnosis, it is also well-known that performance in CPCs in no way reflects a broader measure of competence in a physician’s duties~\cite{ledley1959reasoning}. Furthermore, the act of DDx comprises many other steps that were not scrutinized in this study, including the goal-directed acquisition of information under uncertainty (known to be challenging for AI systems despite recent technical progress in this direction~\cite{hong2023zero, kossen2022active, mackie2023generative}). While based on real-world cases, the clinical pathology case presentation format and input into the model does differ in important ways from how a clinician would evaluate a patient and generate their DDx at the outset of a clinical encounter. For example, while the case reports are created as ``puzzles'' with enough clues that should enable a specialist to reason towards the final diagnosis, it would be challenging to create such a concise, complete and coherent case report at the beginning of a real clinical encounter.

We are therefore very cautious to extrapolate our findings toward any implications about the LLM’s utility as a standalone diagnostic tool. Nevertheless, our controlled evaluation mirrored the findings of other recent works exploring the performance of both LLMs and pre-LLM ``DDx generator’’ in smaller subsets of the NEJM CPCs, which have shown the potential for automated technology to reach the correct DDx in these challenging cases with superior performance to standalone physicians~\cite{kanjee2023accuracy,eriksen2023use,buckley2023accuracy,fritz2022evaluation}. While this represents a step beyond historical attempts at automating DDx in NEJM CPCs, where computerized approaches were deemed overtly unreliable for practical use~\cite{miller1985internist}, such studies also undertook limited consideration of the quality of DDx generated by these automated systems or their role as assistive tools.

Our work extends prior observations by showing not only that the LLM was more likely to arrive at a correct answer or provide the correct answer in a list, but that its DDx were determined by an independent rater to be of higher appropriateness and comprehensiveness than those produced by board certified physicians with access to references and search.  

In our study clinicians had access to both images and tabular data in redacted case reports, while the LLM was only provided with the main body of the text. Though the LLM outperformed the clinicians despite this limitation, it is unknown whether and how much this gap would widen if the LLM had access to the figures and tables. Early evidence suggests the effect might be case/context dependent as other studies have found image access by models to not always improve performance in CPCs ~\cite{buckley2023accuracy}. Furthermore, the integration of multimodal inputs by LLMs is an area of novel research~\cite{li2023llava, tu2023towards}, with a large potential number of data modalities to consider, and little precedent for how information from multiple modalities should be integrated over time for a single case by AI systems. 

The repeated examination of NEJM CPCs by automated systems highlights its promise as a ``benchmark'' for LLM evaluation and development. Benchmarking enables comparisons of models against one another and the ability to evaluate a model’s performance improvements or degradation over time. However, consistency in using CPCs as a scalable benchmark is challenging if reliant upon using human judgement to establish whether a candidate differential matches the ground truth. We utilized an automated approach for comparing our LLM for DDx to a baseline LLM performance (GPT-4). Our estimates varied from that recently published in other studies, despite using the same subset of cases ~\cite{kanjee2023accuracy}. Direct comparisons of different technologies would ideally be conducted by more extensive and blinded human evaluation, including work to ensure reproducibility of the human evaluation protocol, analysis of inter-rater reliability, and the use of metrics reflecting the quality, appropriateness and comprehensiveness of LLM differentials in addition to estimations of accuracy. Our estimates of top-1 and top-10 accuracy, while impressive at close to 30\% and 60\% respectively, highlight noteworthy room for improvement for LLMs, especially for complex cases that are non-pathognomonic (i.e., cases that do not have a sign or symptom that defines a diagnosis). However, as noted above, the CPCs represent ``diagnostic puzzles'' rather than real-world examples of common clinical workflows; and it is therefore important to consider more realistic settings in which LLMs might prove of practical value in medicine.

One such example is the potential for LLMs to assist clinicians in complex diagnoses. Deep learning tools have shown considerable promise in many areas of medicine, but are overwhelmingly used as assistive rather than autonomous tools~\cite{muehlematter2021approval}, given the safety-critical nature of medical practice and the many issues of robustness~\cite{roschewitz2023automatic} and fairness~\cite{obermeyer2019dissecting,seyyed2021underdiagnosis,samorani2022overbooked} seen in deployment. Furthermore, observations of standalone diagnostic accuracy often do not guarantee that an AI tool will improve performance in real-world settings as an assistive tool, and it remains unclear how to optimally integrate AI and human decision-making in medicine~\cite{gaube2021ai}. For LLMs in particular, the known incidence of hallucination/confabulation~\cite{umapathi2023med} might mislead clinicians into inaccurate diagnosis, replicating or even extending findings in other clinical settings that AI systems might actually degrade the performance of clinicians rather than necessarily improving outcomes.

This highlights the importance of focused study of LLMs in assistive scenarios. We explored this specifically in NEJM CPCs and found that the proposed LLM for DDx, increased the number of appropriate DDx produced by a clinician when used as an assistive tool in addition to overall top-N accuracy, suggesting that the LLM’s primary assistive potential may be due to making the scope of DDx more complete. Given the potential for misleading information to arise from AI systems, including in convincing dialogue, clinicians must appreciate the fundamental limitations of these models and not lose sight of their primacy in the provider-patient relationship and their ultimate authority and responsibility for the diagnostic and therapeutic management of their patients. Such thoughtful and effective LLM use should not be unintuitive to most clinicians. Aiding the diagnostic process could reasonably occur in an emergency room upon presentation (during potentially time-sensitive moments), upon admission to the medical ward, or by a consulting service after a patient has been admitted or in outpatient clinics. Our findings suggest that onward research should more rigorously explore how LLMs augment clinicians’ DDx in many such specific scenarios, where the risks and benefits might vary.

Despite being a novel tool, the use of the LLM did not seem to add inefficiency or increase the amount of time spent on solving each CPC compared to the use of Search or other conventional information. This suggests that the conversational interface was unobtrusive and intuitive. Consistent with this, the interviewed clinicians all described it as ``easy'' to use, and were positive about the use and implications of the LLM interface. Enhancing efficiency while maintaining or improving quality are generally accepted goals of improving health care delivery, alongside improving provider experience~\cite{sikka2015quadruple}, and our study showed significant potential in this regard, as clinicians also reported feeling more confident in their DDx lists after using the model. ``That is where the search really became difficult, I didn't know what to put in to narrow down my search that is going to help me narrow down my differential.''
However, there are numerous human factors, social elements, and other complex considerations in these use cases, and it is critical to ensure efforts are made to avoid inequities in access to avoid exacerbating existing health disparities.  

Clinicians frequently expressed excitement about using the LLM but were also aware of the shortcomings of language models and had concerns about confabulations in particular if used by individuals not trained or instructed to avoid such questions. However, our work did not explore many other important aspects of human-AI interaction, which require further study in safety-critical settings such as this. For example, we did not explore the extent to which clinicians trusted the outputs of the model or their understanding of its training and limitations, or undertake focused ``onboarding'' or training in its use, which are all known to be important modifiers of the benefits derived by clinicians from AI assistants~\cite{cai2019hello}. The CPC challenges themselves do not enable a rigorous exploration of the possible impacts of AI assistance on health equity and fairness; a further study of how these aspects of clinicians’ DDx is impacted by LLM assistance is needed. While AI systems are known to be able to express uncertainty~\cite{yin2023large} and defer appropriately to clinicians~\cite{dvijotham2023enhancing}, which might significantly improve the balance between trust and skepticism needed for effective AI assistance in medicine.  Qualitative feedback suggested that there remains room for targeted improvement of LLMs as assistive diagnostic tools, with one clinician noting that \textit{``It was most helpful for simpler cases that were specific keywords or pathognomonic signs.''} (C3) but for more complex cases it still tended to draw conclusions from isolated symptoms rather than viewing the case holistically. The assistive effect of these LLMs could potentially `upskill’ clinical providers, particularly in enabling them to broaden and enhance the quality of their DDx. As corroborated via our clinician interviews after their experience with the LLM, such upskilling could be relevant for education or training purposes to support providers across a skill continuum ranging from trainees to attending providers. The upskilling capabilities could also extend to locations where specialist medical training is less common (e.g., in lower and middle income countries [LMIC]). However, our findings may not generalise to these scenarios, given that we utilized a pool of twenty clinicians with a mean experience of 11.5 years. This may not adequately represent the diverse set of users seeking to benefit from LLMs as a diagnostic aid. Further studies are warranted in an array of more realistic clinical scenarios, with a wider range of clinical users that might range from medical students to allied health professionals. The underlying mechanism of assistance also merits further study and might be an important factor for optimising the assistive effect of LLMs, particularly given their conversational and interactive nature. For example, our study did not explore the impact of LLMs on the clinical reasoning process.

\section{Limitations}

There are limitations to this evaluation. While based on real-world cases, the clinical pathology case presentation format and input into the model does differ in important ways from how a clinician would evaluate a patient and generate their differential diagnosis at the outset of a clinical encounter. The case reports are created as ``puzzles'' with enough clues that should enable a specialist to reason towards the final diagnosis. At the beginning of a clinician encounter, it would be challenging to create such a concise, complete and coherent case report. Case reports in the NEJM style would not be available when at intake. Similarly, these cases were selected to represent challenging cases instead of common conditions (i.e., `zebras' as opposed to `horses' in clinical parlance). As such, our evaluation does not directly indicate the results of or suggest that clinicians should leverage the assistive capabilities of an LLM for typical cases seen on a daily basis.

In terms of modalities, the case reports include both images and tables. The clinicians had access to these in the redacted case reports. However, the LLM only had access to the main body of the text. Though the LLM for DDx outperformed the clinicians despite this limitation, it is unknown whether and how much this gap would widen if the LLM had access to the figures and tables. Early evidence suggests the effect might be case/context dependent~\cite{buckley2023accuracy}.

The study highlighted some weaknesses of the existing LLM. Specifically, one clinician highlighted that \textit{``It was most helpful for simpler cases that were specific keywords or pathognomonic signs.''} (C3) but that for more complex cases it still tended to draw conclusions from isolated symptoms rather than viewing the case holistically. Considering the significance of assessing challenging cases, the NEJM CPC case reports will likely serve as a useful dataset for continued benchmarking as LLMs improve in performance.

\section{Conclusion}
\label{sec:conclusion}

Generating a DDx is a critical step in clinical case management, and the capabilities of LLMs present new opportunities for assistive tooling to help with this task. Our randomized study showed that the LLM for DDx was a helpful AI tool for DDx generation for generalist clinicians. Clinician participants indicated utility for learning and education, and additional work is needed to understand suitability for clinical settings.

\vspace{12pt}
\subsubsection*{Acknowledgments}
This project was an extensive collaboration between many teams at Google Research and Google DeepMind. 
We thank Ayush Jain, Rory Sayres, Sami Lachgar, Lauren Winer, Maggie Shiels, Brett Harfield, Si Wai Man, Preeti Singh, Annisah Um'rani, Bradley Green, and Philip Mansfield for their valuable insights and feedback during our research. We are also grateful to Micheal Howell, Meredith Morris, Celeste Grade, Karen DeSalvo, Zoubin Ghahramani, James Manyika, and Jeff Dean for their support during the course of this project. Finally, we extend our sincere thanks to the Massachusetts Medical Society group for the support and partnership.

\subsubsection*{Competing interests}
This study was funded by Alphabet Inc and/or a subsidiary thereof (‘Alphabet’). All authors are employees of Alphabet and may own stock as part of the standard compensation.

\newpage
\setlength\bibitemsep{3pt}
\printbibliography
\balance
\clearpage


\end{refsection}
\end{document}